\documentclass[manuscript]{aastex}

\shorttitle{Development of SQUEAN}
\shortauthors{Kim et al.}

\begin{document}

\title{Development of SED Camera for Quasars in Early Universe (SQUEAN)}

\author{Sanghyuk Kim$^1$, Yiseul Jeon$^2$, Hye-In Lee$^1$, Woojin Park$^1$, Tae-Geun Ji$^1$, Minhee Hyun$^{2, 3}$, Changsu Choi$^{2, 3}$, Myungshin Im$^{2, 3}$, and Soojong Pak$^{1, \star}$}
\affil{$^1$School of Space Research and Institute of Natural Sciences, Kyung Hee University, 1732 Deogyeong-daero, Giheung-gu, Yongin-si, Gyeonggi-do 446-701, Korea}
\affil{$^2$Astronomy Program, Department of Physics \& Astronomy, Seoul National University, 1 Gwanak-ro, Gwanak-gu, Seoul 151-742, Korea}
\affil{$^3$Center for the Exploration of the Origin of the Universe (CEOU), Astronomy Program, Department of Physics \& Astronomy, Seoul National University, 1 Gwanak-ro, Gwanak-gu, Seoul 151-742, Korea}

\begin{abstract}
We describe the characteristics and performance of a camera system, Spectral energy distribution Camera for Quasars in Early Universe (SQUEAN).
It was developed to measure SEDs of high redshift quasar candidates (z $\ga$ 5) and other targets, e.g., young stellar objects, supernovae, and gamma-ray bursts, and to trace the time variability of SEDs of objects such as active galactic nuclei (AGNs).
SQUEAN consists of  an on-axis focal plane camera module, an auto-guiding system, and mechanical supporting structures.
The science camera module is composed of a focal reducer, a customizable filter wheel, and a CCD camera on the focal plane.
The filter wheel uses filter cartridges that can house filters with different shapes and sizes, enabling the filter wheel to hold twenty filters of 50 mm $\times$ 50 mm size, ten filters of 86 mm $\times$ 86 mm size, or many other combinations.
The initial filter mask was applied to calibrate the filter wheel with high accuracy and we verified that the filter position is repeatable at much less than one pixel accuracy.
We installed and tested 50 nm medium bandwidth filters of 600 -- 1,050 nm and other filters at the commissioning observation in 2015 February.
We found that SQUEAN can reach limiting magnitudes of 23.3 - 25.3 AB mag at 5-$\sigma$ in a 1-hour total integration time.
\end{abstract}

\section{Introduction}
Slit-less spectroscopy using a prism or a grism is an effective technique for wide-field survey to obtain multi-object, low-resolution spectra \citep{Cram1989, Hickson1994}.
However, this method has the disadvantages of a bright sky background and overlapping spectra in crowded fields.
Multi-object spectrographs employing optical fibres or multiple slits can reduce the sky background, but have a limitation that a only relatively small number of objects can be simultaneously observed \citep{Hickson1994}.
Recently, multi-color broad-band photometry methods have been applied to classify objects into galaxies, quasars, and stars of various kinds and to estimate their redshifts \citep{Drink1988, Koo1985, Loh1986, Rich2001, Rich2009, Wolf2001a, Wolf2001b}.
However, the limited number of wavelength bands and wide filter widths allow us to do only a rough classification of objects and achieve redshift estimates of a few \% or worse \citep{Kang2015, Kim2015b, Lee2015, Wolf2001a, Wolf2001b}.

In order to overcome the above short-comings of broad-band imaging or multi-object spectroscopy, \citet{Hickson1994} proposed a technique, intermediate between the extremes of multi-color and multi-slit observations, that involves imaging through sequential series of medium or narrow-band filters.
One clear advantage of medium-band use can be highlighted as follows.
The medium-band imaging can trace the continuum spectral energy distribution (SED) shapes of galaxies and quasars much better than broad-band observations, which improves object classification and photometric redhshift measurements.
For example, Classifying Objects by Medium-Band Observations (COMBO-17; \citealt{Wolf2003a, Wolf2003b}) used seventeen medium-band filters and produced a catalog of galaxies and quasars with photometric redshift errors of $dz$/$(1+z)$ $\simeq$ 0.01 for bright galaxies \citep{Wolf2008}.
Following the same philosophy, the Advanced Large, Homogeneous Area Medium-Band Redshift Astronomical (ALHAMBRA) survey was performed using medium-band optical filters from 350 nm to 970 nm and standard broad-band filters of $J$, $H$, and $K$ in the near-Infrared (NIR) wavelength range \citep{Apa2010, Mat2013, Mol2008}.
The photometric redshift errors can be improved even more to the 0.3\% level by using more medium-band filters \citep{Ben2014}.
The Cosmic Evolution Survey (COSMOS) used 30 medium-band and broad-band filters, and they also reported that the use of many filters enables them to achieve photometric redshift accuracy of the 0.7\% level for galaxies \citep{Ilbert2009}.
High precision photometric redshifts can be used for cosmological study of large scale structures.
Medium-band dataset even allows measurements of photometric redshifts to the 1\% level for quasars \citep{Jeon2016, Wolf2008}.
Because of these advantages, several surveys adopted this approach in the last decade \citep{Ben2009a, Ben2009b, Ben2012, Davis2003, Hickson1998, Tan2005, Wolf2001a, Wolf2001b}.

Another advantage of medium-band filter observation is its potential for tracing broad spectral features such as broad emission lines of quasars.
Quasars, regardless of their redshifts, typically have broad emission lines with full width at half maximum (FWHM) of a few thousand km s$^{-1}$ or more (e.g., \citealt{Jun2015, Kim2015a}) which matches nicely the width of medium-band filters.
For that reason, flux measurements of emission lines and continuum are possible with medium-band observation, and this enables time series study of quasars, such as the reverberation mapping (e.g., \citealt{Kaspi2000}) to measure black hole masses in quasars (see \citealt{Fau2016} for an example of reverberation mapping using broad-band photometry).
Strong narrow emission lines of star forming galaxies can be identified for measuring their redshifts and star formation rates too.

From 2010 to 2014, we have been using Camera for Quasars in Early Universe (CQUEAN), an imaging camera with an enhanced sensitivity with a wavelength of 0.7 - 1.1 $\micron$ with a field of view (FOV) of $4\farcm7 \times 4\farcm7$ at the 2.1 m Otto Struve Telescope of the McDonald Observatory to perform follow-up imaging observations of high redshift quasar candidates \citep{Kim2011, Lim2013, Park2012}.
CQUEAN consists of a science camera, broad-band filters, a focal reducer to increase the FOV, and a guide camera system with its in-house control software.
In order to implement medium-band filter observation capabilities, we upgraded CQUEAN to SED CQUEAN (SQUEAN) in 2015 February.
The primary changes from the CQUEAN to SQUEAN are a new filter wheel system that houses a suite of medium-band filters, a new auto guiding system, and new software packages to support new hardware operation.
SQUEAN has the virtues of high sensitivity over a wide wavelength range, simple structure with twenty exchangeable filters and a wide compatibility with variable camera systems.
With this upgrade, we are now performing low resolution spectroscopy observation (i.e., medium-band imaging) of faint quasar candidates at $z > 5$ ($z \sim 23$ AB mag) that have been identified in the Infrared Medium-deep Survey (IMS; e.g., see \citealt{Kim2015a}) and medium-band reverberation mapping of lower redshift quasars.
The use of SQUEAN allows us to identify and measure redshifts of faint quasars at a moderate exposure time using 2-m class telescopes, which, otherwise, requires  time-expensive spectroscopy with 8-m class telescopes.

In this paper, we present an overview of SQUEAN and its performance based on test observations.
In Section 2, we describe the design, structure analysis, control software, and performance of SQUEAN.
We report the initialization of a filter wheel and the stability test results in Section 3.
Section 4 details the system performance.
Finally, a summary is presented in Section 5.

\section{Instrument Description}

\subsection{System Architecture}
\begin{figure}
\plotone{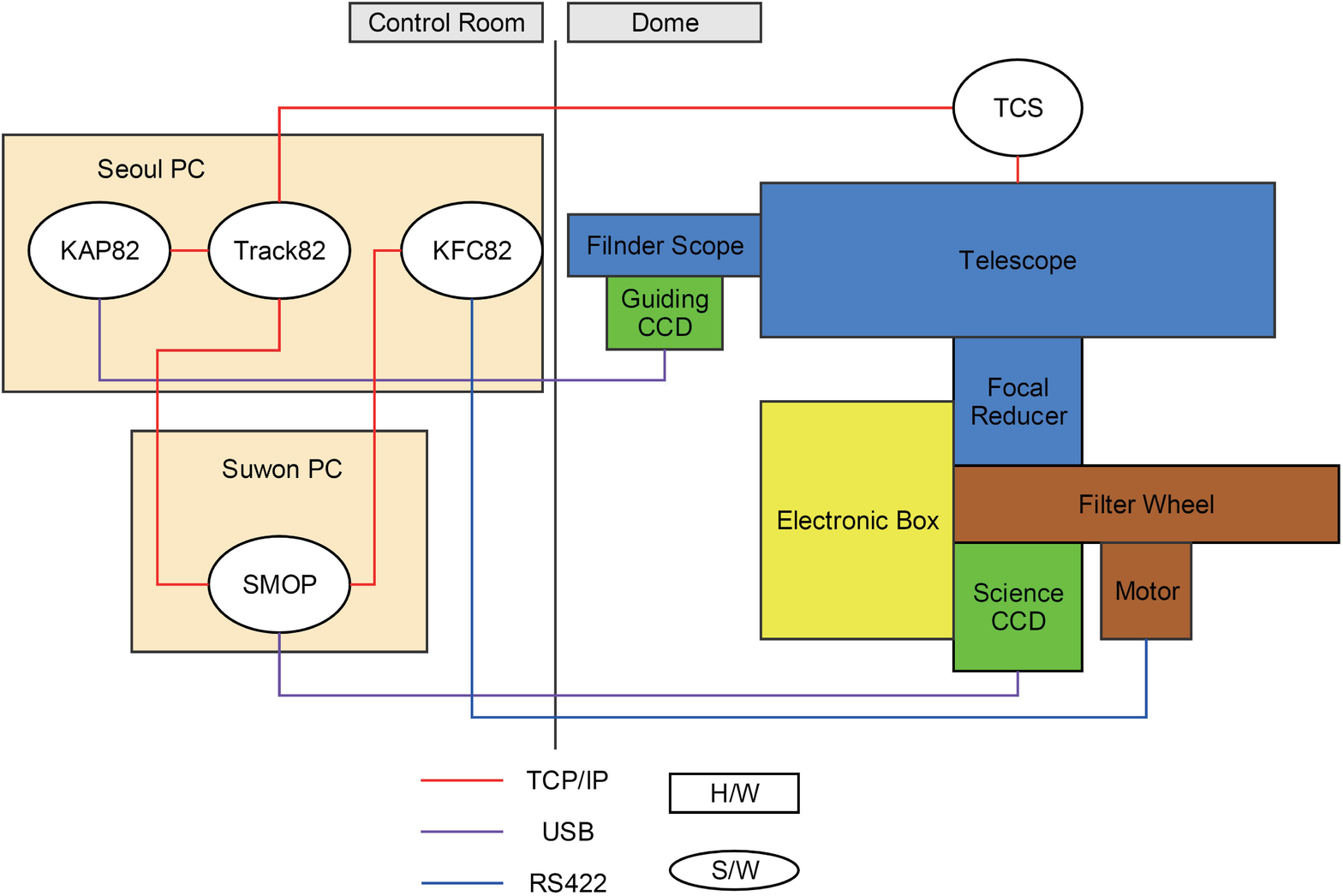}
\caption{System architecture of SQUEAN.\label{System}}
\end{figure}

\begin{figure}
\plotone{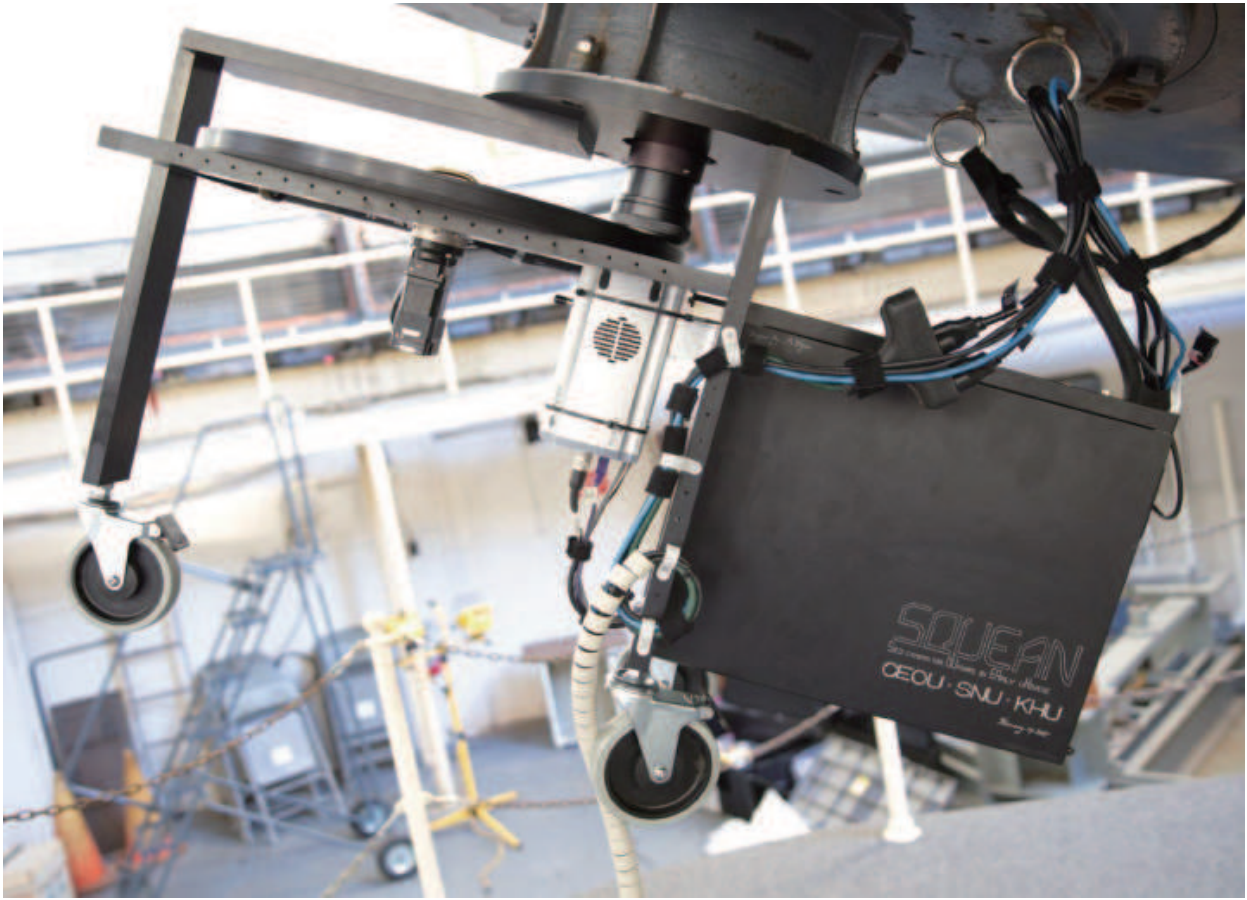}
\caption{SQUEAN attached to the Cassegrain focus of the 2.1 m telescope. The black box on the right is the electronics box and the small silver one is the CCD camera. The large black disk on the left is the filter wheel. The step motor and the planetary reducer are assembled at the bottom of the filter wheel.\label{SQUEAN}}
\end{figure}

Figure \ref{System} shows the system architecture of SQUEAN, which consists of the science, the auto guiding, and the supporting parts.
Figure 2 shows SQUEAN on the Cassegrain focus of the 2.1 m telescope (see Figure \ref{SQUEAN}).
In the science part, the focal reducer and the CCD camera are from the components of CQUEAN \citep{Kim2011, Lim2013, Park2012}, but we designed and fabricated a new filter wheel that can house twenty medium-band filters.
The new auto-guiding system is attached to the finder scope of the main telescope (see the description of the preliminary system in \citealt{Choi2015}), since the new, large filter wheel obstructs the off-axis light that were previously used in the CQUEAN auto-guiding system \citep{Kim2011}.
It consists of an optical CCD camera and a filter wheel that contains Johnson $BVRI$ filters and a block filter.
Using this filter wheel and a guide CCD camera, the auto-guiding system can also be operated independently for its own science programs.

\subsection{Filters}
SQUEAN includes nine medium-band filters that cover wavelengths from 600 to 1,050 nm, each having 50 nm width, spaced at 50 nm apart from the adjacent filter.
Photometry with these medium-band filters corresponds to spectroscopy with a resolution of R $\sim$ 15, and can trace SEDs of various objects better than broad-band photometry.
The usefulness of the medium-band filters for the selection and redshift measurement of high redshift quasars has been demonstrated in \citet{Jeon2016}.

The medium-band filters are all commercial off-the-shelf products from the Edmund Optics Inc.
We name the filters using a combination of the letter ``$m$'', the initial of ``medium'', and their central wavelengths in nm, e.g., $m625$, $m675$, and $m1025$.
Diameters and bandwidths of all the filters are 50 mm and 50 nm, respectively, except for filters with the central wavelengths at 925 nm due to the product availability.
For the 925 nm central wavelength filter, we use $m925s$ and $m925n$ filters.
The $m925s$ filter has a diameter of 25 mm and a bandwidth of 50 nm.
The $m925n$ filter has a diameter of 50 mm and a bandwidth of 25 nm.

In addition to the series of the medium-band filters, we include broad-band filters, i.e. CQUEAN filters ($g$, $r$, $i$, $z$, $Y$, $iz$ and $is$) and Johnson-Cousin $BVRI$ filters.
The detailed specification of CQUEAN filters are given in \citet{Park2012}.
The Johnson-Cousin $BVRI$ filters are 50 $\times$ 50 mm, in size and manufactured by Astrodon\footnote{http://dg-imaging.astrodon.com.}.
Note that each filter is mounted on its own cartridge.
This makes the installation of the cartridge simple and we can exchange filters during observation.
Also, it enables the filter wheel to accept filters with various sizes and shapes for customized usage of the system.

\begin{figure}
\plotone{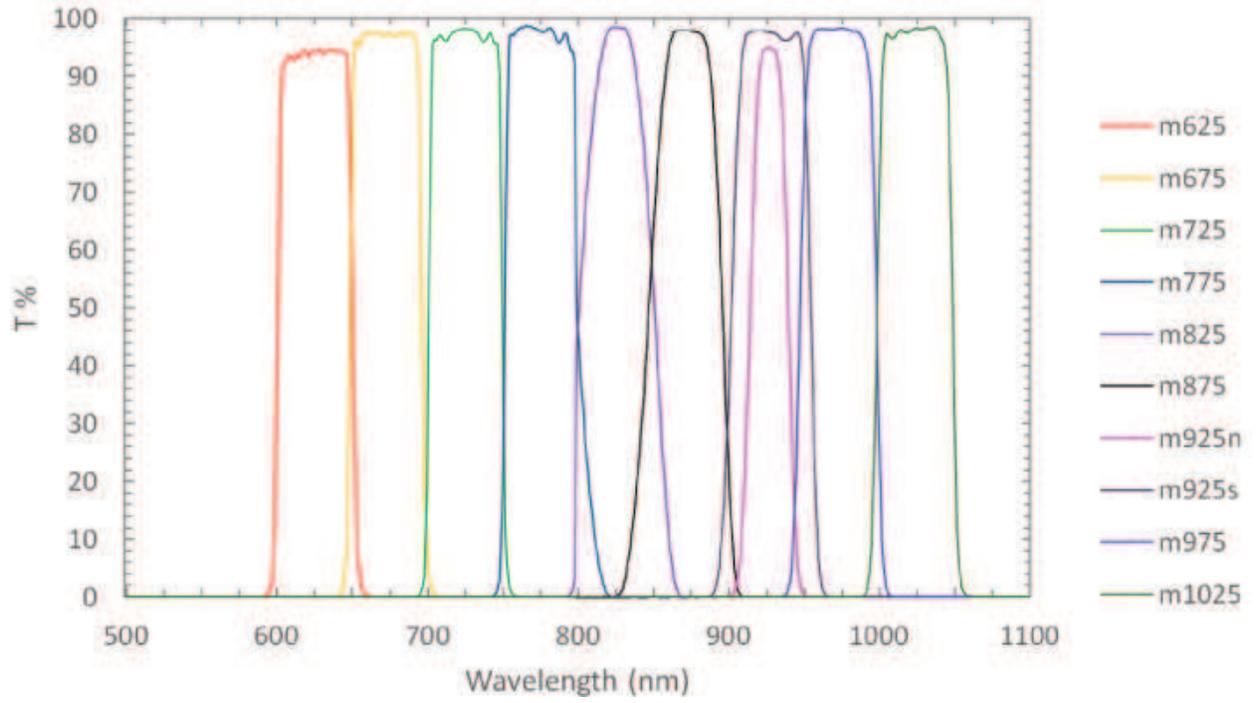}
\caption{Transmission curves of the medium-band filters for SQUEAN.\label{TransM}}
\end{figure}

\begin{figure}
\plotone{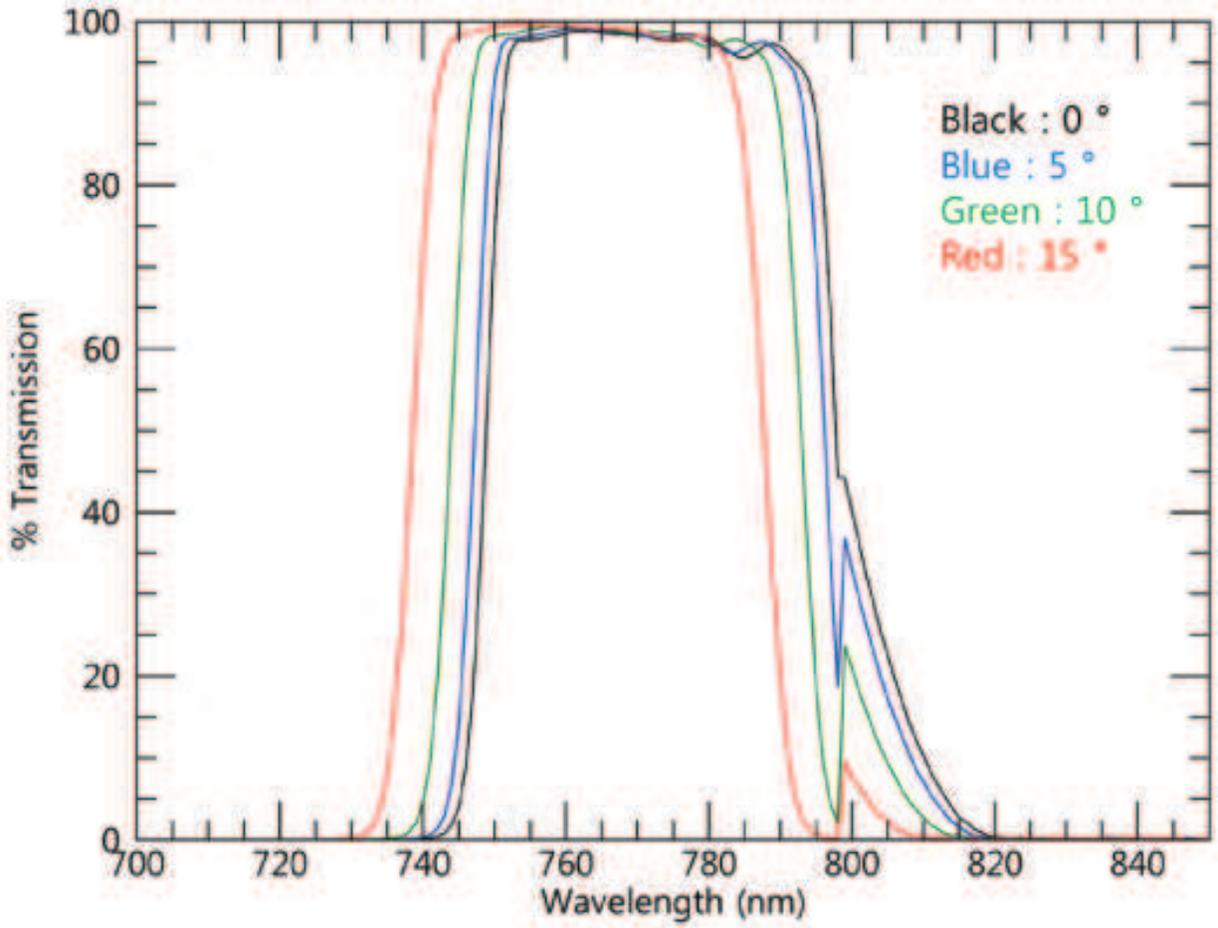}
\caption{Transmission curves of the $m750$ filter for various incident angles. The bumps on the right wings of the plot are artificial effects caused by the light source change of the spectrophotometer.\label{Shift}}
\end{figure}

Figure \ref{TransM} shows the transmission curves of the medium band filters, which were measured using a spectrophotometer, Cary 400 (Varian Inc.\footnote{https://www.agilent.com.}).
As shown in Figure \ref{Shift}, we also checked the transmission curves of the filters for the incident angles of 5$\degr$, 10$\degr$, and 15$\degr$.
The transmission curves are shifted to a shorter wavelength when the incident angle is increased.
The shift of the transmission curve is less than 2 nm for the incident angle of 5$\degr$ and increases in proportion to the incident angle.
But the maximum incident angle deviation of SQUEAN optics is 4.3$\degr$ and the wavelength shift is negligible for SQUEAN's 50 nm width filters.

\begin{figure}
\plotone{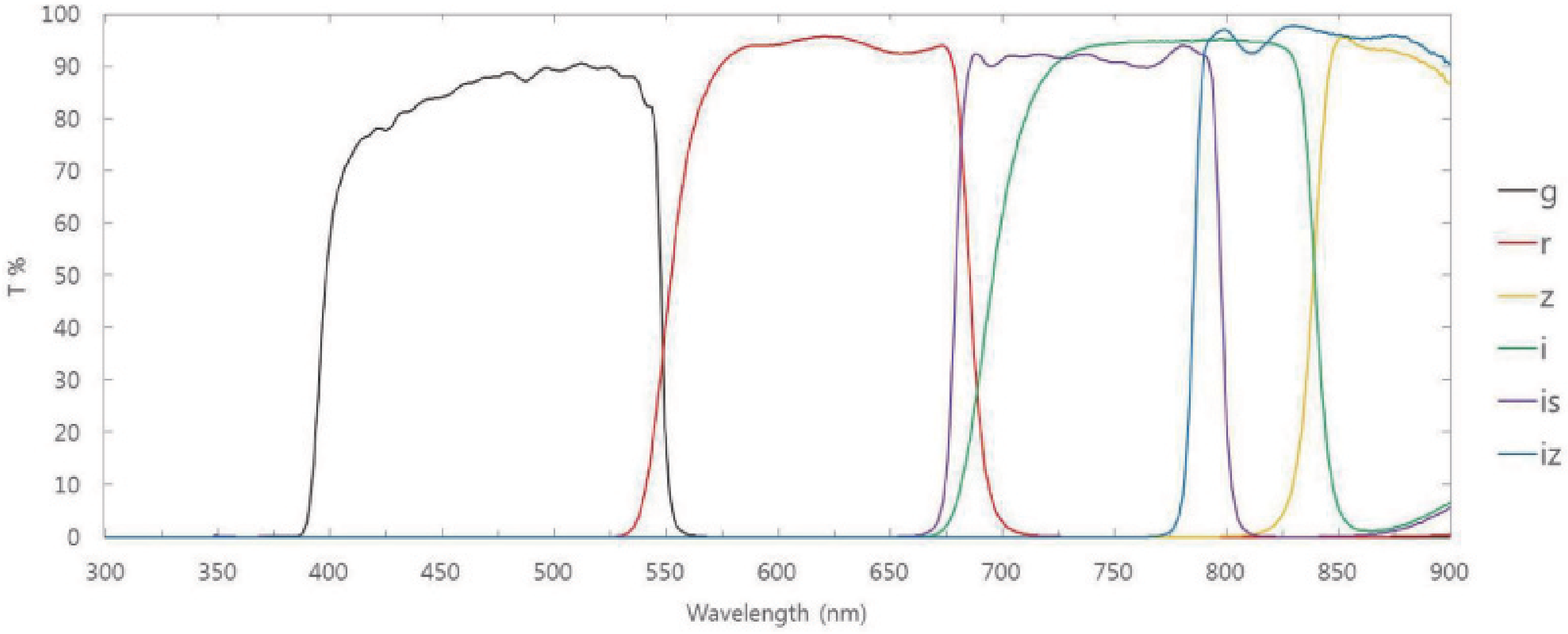}
\plotone{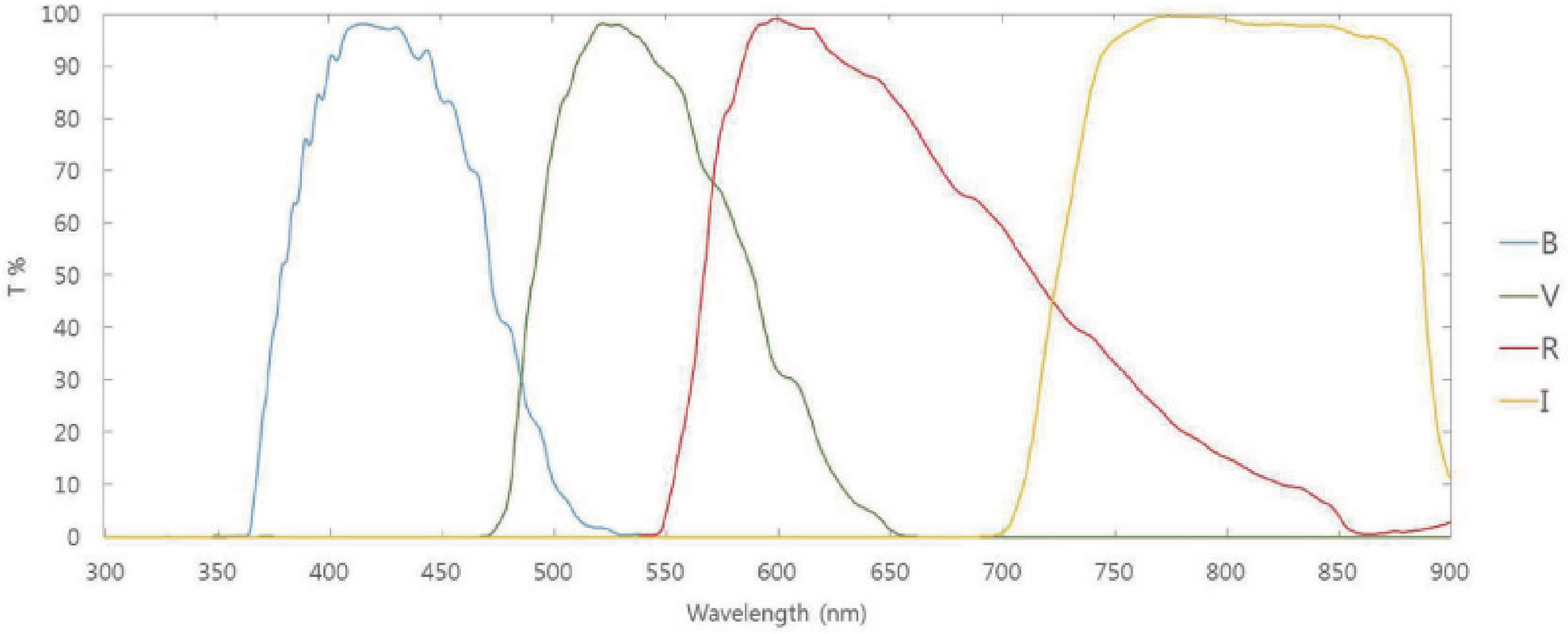}
\caption{Transmission curves of the broad-band filters for SQUEAN. The upper image shows the transmission curves of CQUEAN filters and the bottom is for the $BVRI$ filters.\label{TransB}}
\end{figure}

Figure \ref{TransB} shows the transmission curves of the broad band filters, which were measured using a spectrophotometer Cary 100 (Varian Inc.$^2$) in McDonald observatory.
Because of the measurement limits of the machine, the available transmission curves of $iz$ and $z$ filters are cut at 900 nm.

\subsection{Filter Wheel}
\begin{figure}
\plotone{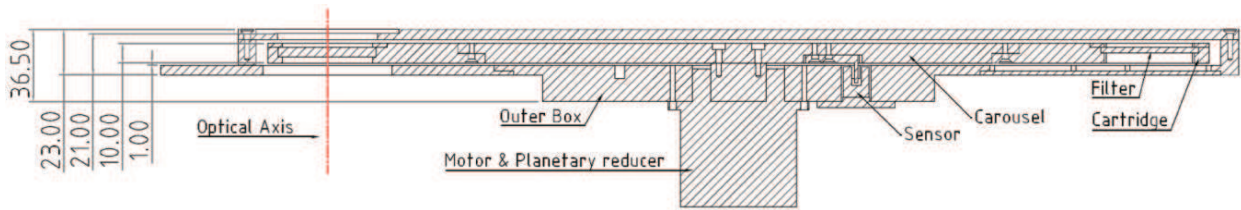}
\caption{Cross-section image of the filter wheel.\label{FW}}
\end{figure}

The primary improvement from CQUEAN to SQUEAN is a new large filter wheel that can house twenty filters with a square size of 50 mm $\times$ 50 mm.
Figure \ref{FW} indicates the cross-section of the filter wheel.
In order to satisfy the flange focal length of the focal reducer, the thickness of the filter wheel is limited to within 21 mm.
The thicknesses of the outer box and the cover of the filter wheel are 5 mm and 3 mm, respectively.
The gaps from the carousel to the outer box and the cover are 1 mm.
Because the thickness of the cartridge cover is 1 mm, the maximum thickness of the filter is limited to 10 mm.
The mechanical structure is modularized into three parts: (1) the filter wheel box, which is comprised of the outer box, the cover for a focal reducer, a CCD adaptor and a window to replace filter cartridges and filters; (2) the control electronics, i.e., a step motor, a planetary reducer, an optical sensor and a chopper; and (3) the rotating disk module, i.e., a carousel, interchangeable cartridges, and filters.
The filter wheel can be attached to various CCD cameras and telescopes by simply replacing the adapting plates.
In addition, the exchange of filter cartridges is possible without disassembling the filter wheel from the telescopes and cameras.

\begin{figure}
\plotone{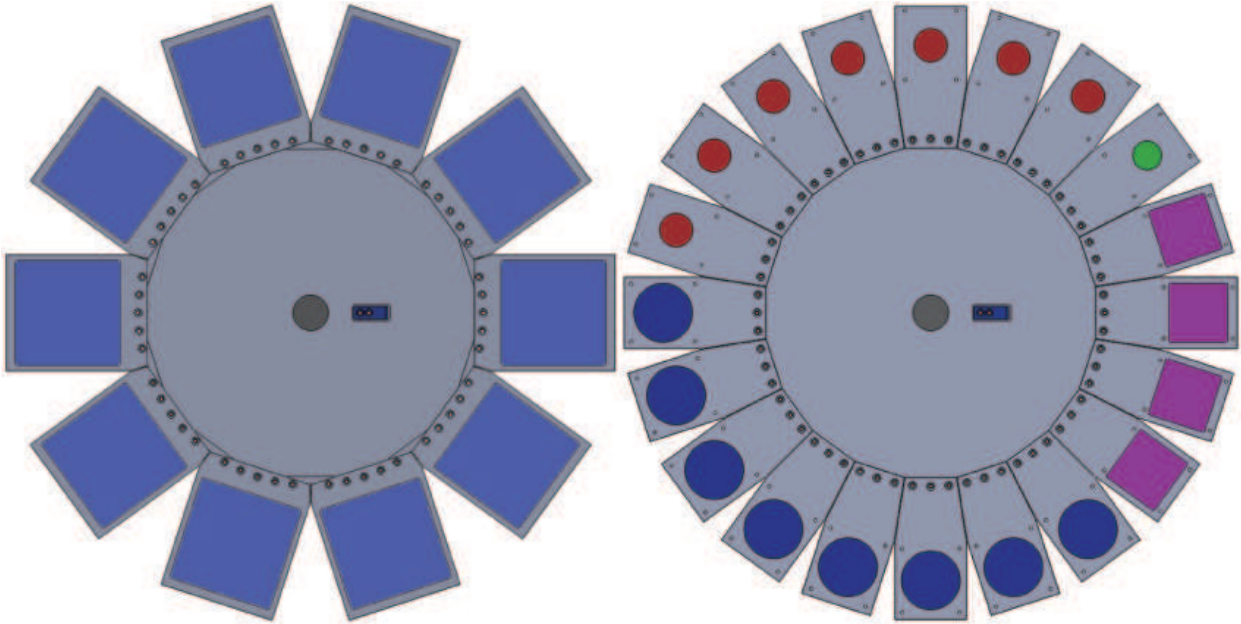}
\caption{Carousel with different types of cartridges. In the left image, large size cartridges are installed. The right image shows the normal design concept, which can assemble twenty filters of less than 7 mm thickness and 50 mm $\times$ 50 mm size.\label{Caro}}
\end{figure}

As shown in Figure \ref{Caro}, the carousel is designed to install twenty normal or ten large interchangeable cartridges.
The normal cartridge can be assembled with filters of less than 10 mm thickness and a 50 mm $\times$ 50 mm size.
The largest cartridge can be associated with the filter size up to 86 mm $\times$ 86 mm.

\begin{figure}
\plotone{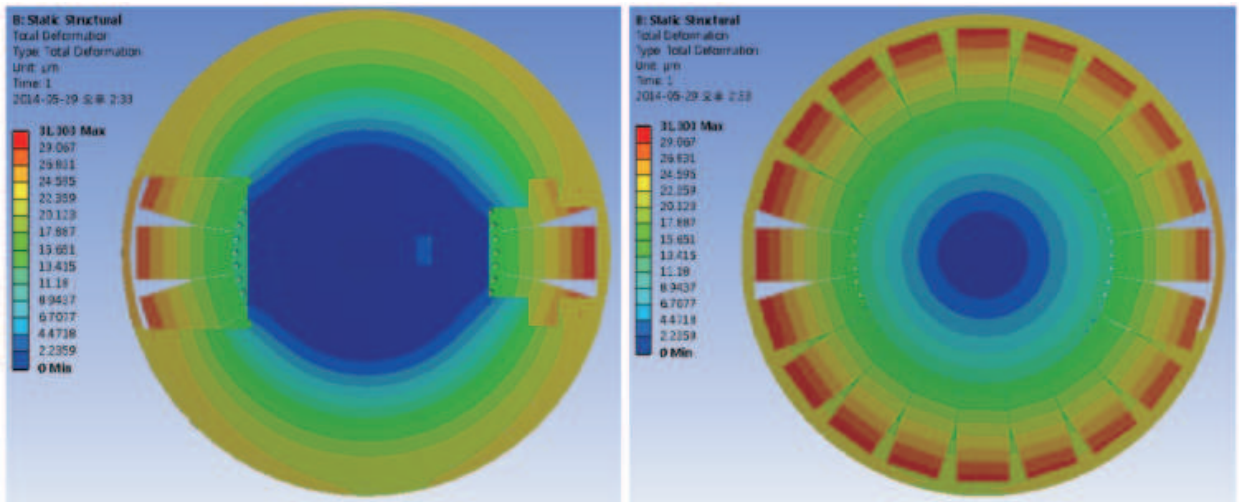}
\caption{FEA results of the filter wheel flexure. The maximum flexure of the filter wheel is less than 32 $\micron$.\label{FEA}}
\end{figure}

The filter wheel and its components are designed densely but very thinly by the thickness restriction.
The heavy components such as filters are assembled at the edge of the carousel.
The flexures of the filter wheel and the carousel cause the tilt of the optical axis and degenerate the optical performance.
To check the flexure of the filter wheel components due to self-weight, we performed finite element analysis (FEA).
The critical deformed components are the carousel and the outer box.
With a thickness of 10 mm, the carousel axis is attached to the top of a planetary reducer and its outer surrounding edge holds many cartridges and filters.
The diameter of the outer box is 520 mm and the thinnest thickness is 5 mm.
We place the carousel 1 mm above the outer box and 2 mm below the cover to prevent mechanical collision.
Figure \ref{FEA} shows the analysis results, and the estimated maximum flexure of the filter wheel is 32 $\micron$, which is much smaller than the gap between the carousel and the outer box.
This error corresponds to the tilt angle of 0.$\arcmin$5 in the optical axis, which does not affect the optical performance of SQUEAN.

Instead of a conventional bearing and chain system, we selected the step motor and the planetary reducer to drive the filter wheel.
The step motor includes an encoder for accurate positioning of the filters.
This selection allows a simple and efficient mechanical design.
We also used the optical sensor and the chopper in lieu of mechanical sensing for initial positioning of the filters.
To facilitate replacement of the sensor system in case of malfunction, we isolated the sensor box from the outer box.

\subsection{Control Software}
As shown in Figure \ref{System}, the new control system of SQUEAN is composed of four software packages that are operated by two computers.
The Suwon PC controls the SQUEAN Main Observation software Package (SMOP) and the Seoul PC manages the KHU Filter wheel Control software package (KFC82), KHU Auto-guiding software Package for McDonald 82 inch Telescope (KAP82), and Track82 which is the telescope control software package for the 2.1 m telescope.
Among these, SMOP, KFC82, and KAP82 are new systems and only the Track 82 software was used in CQUEAN.

\begin{figure}
\plotone{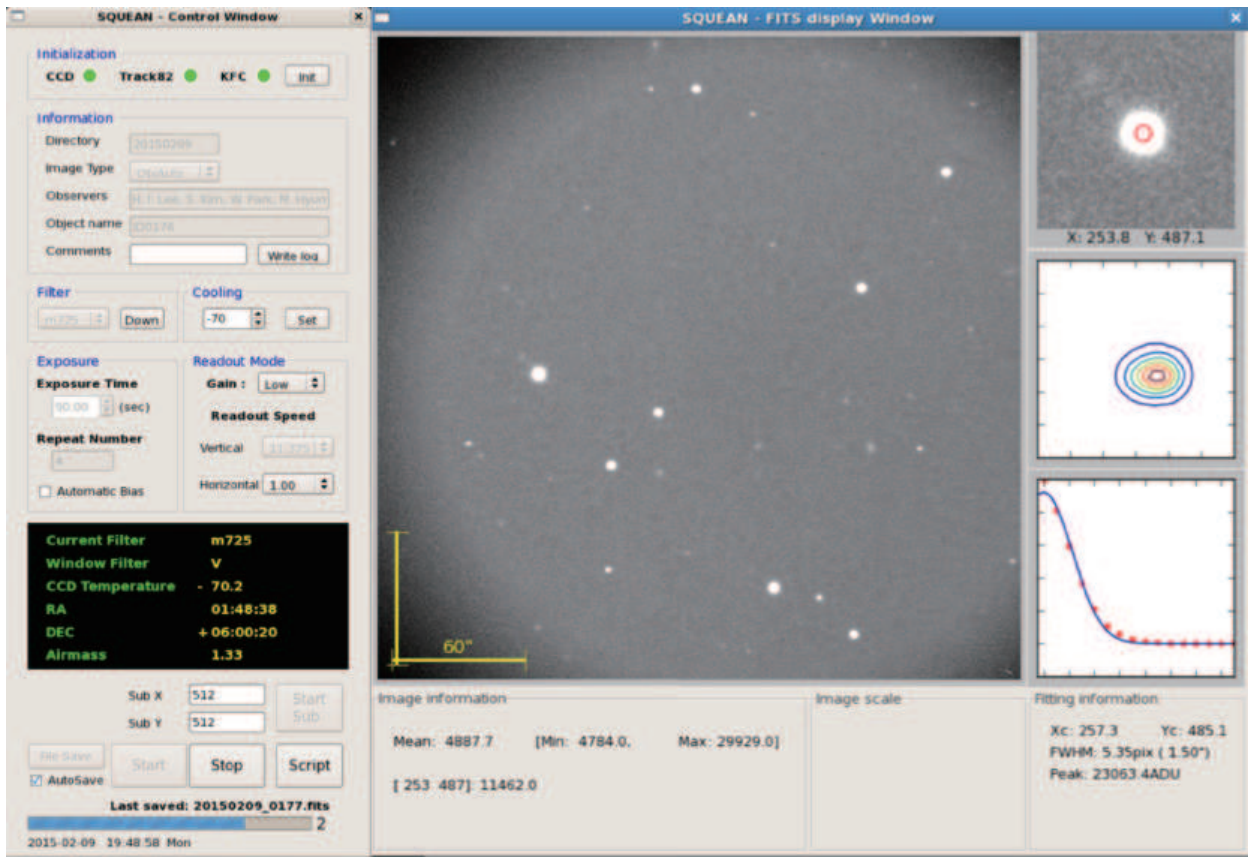}
\caption{Screenshot of SMOP running in CentOS. The left window is the control panel and the right one is a quick look display of the target.\label{SMOP}}
\end{figure}

Figure \ref{SMOP} shows two windows of SMOP.
The left window is the control panel and the right is the quick look display of a target.
SMOP is a Python-based graphic user interface (GUI) program and it runs on CentOS.
It is connected to other control software packages such as KAP82, KFC82, and the Andor camera control software by socket communication.
KAP82 replaces the agdr (the old CQUEAN auto-guiding software), and it operates on the Seoul PC.
Originally, KAP82 was written in Python on a LINUX platform, which was originally named CAP, as in \citet{Choi2015}, but then it was redeveloped with Visual \verb!C++! on Microsoft Windows.
It was developed on the basis of the auto-guiding software package for the Immersion GRating INfrared Spectrograph (IGRINS) slit viewing camera \citep{Kwon2012, Park2014}.
The GUI of KAP82 was designed to be similar to that of the previous auto-guiding software package to minimize confusion.

KFC82 is a GUI program that is written in Visual \verb!C++! and also runs on the Microsoft Windows operating system.
The filter wheel can be controlled by two GUIs: the KFC control panel and SMOP control panel.
The KFC82 control panel provides functions for the calibration, the initialization, and the test of the filter wheel.
In this panel, we can control and check the status of the filter wheel.
We can also edit the filter list on this panel.
During observation, KFC82 can be controlled by the the main control panel of SMOP, so the observation sequence script of SMOP can automatically select the filters.

\subsection{Operation Scenario}
\begin{figure}
\plotone{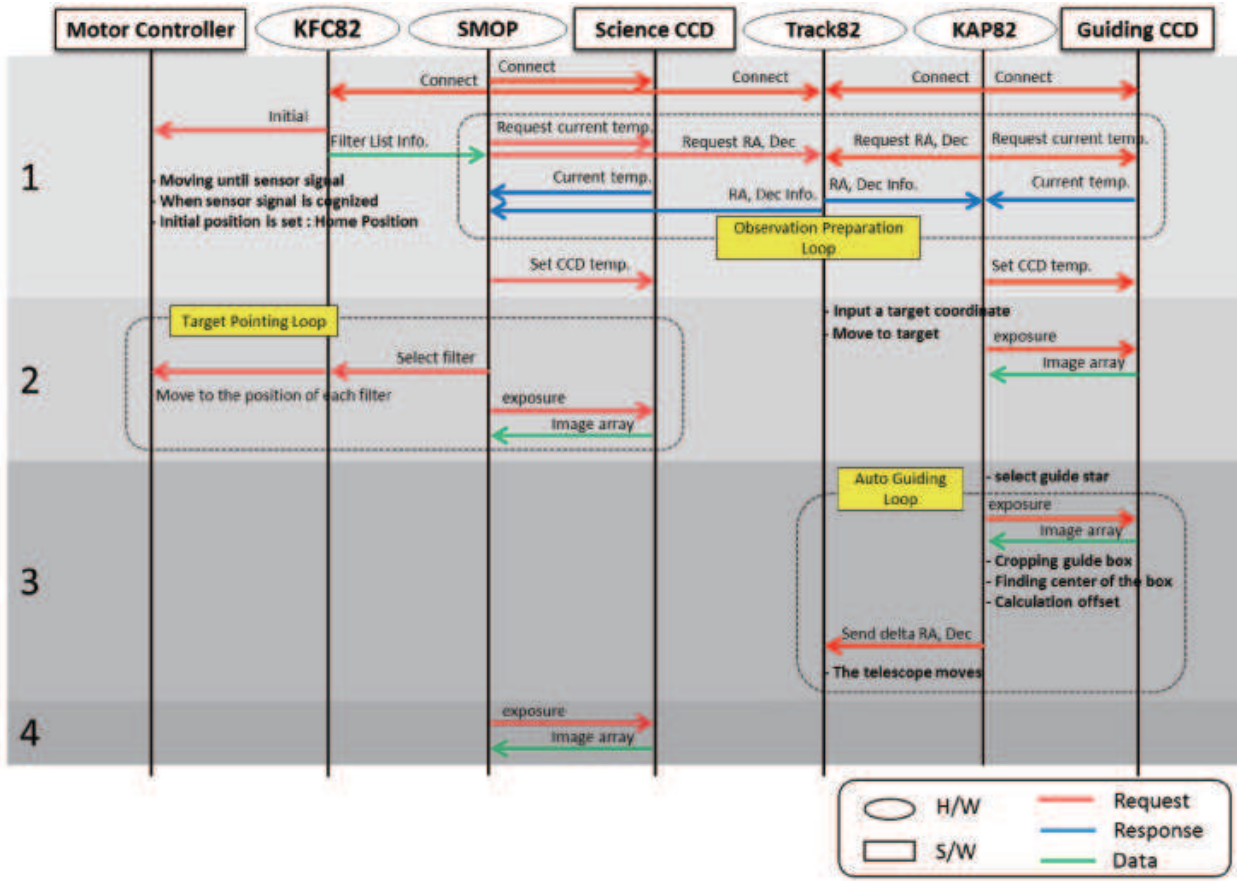}
\caption{Sequence map of the SQUEAN software system. The numbers on the left side represent the operation phases for observation.\label{Sequence}}
\end{figure}

Figure \ref{Sequence} describes the software sequence map.
The observation process is conducted in four phases:\\
(1) The first is the observation preparing phase.
SMOP connects with science CCD, KFC82, and Track82, while KAP82 is linked to Track82 and guiding CCD.
Both SMOP and KAP82 receive the right ascension (R.A.) and declination (Dec) information from Track82 and start the CCD cooling.
KFC82 also starts the initialization of the filter wheel.\\
(2) The second is a target pointing phase.
An observer types the target coordinates to Track82 to move the telescope to the target.
Track82 transfers the coordinate information by the request of SMOP.
After the target is acquired, the observer selects the filter and SMOP takes a test image to confirm the target.\\
(3) In the third phase, an observer selects a guide-star through KAP82, and KAP82 starts guiding on the star.
KAP82 continuously communicates with Track82 to check the R.A, DEC information and its offset.\\
(4) In the fourth phase, the SMOP starts the observation with an appropriate exposure time.

\section{Calibration and Stability Test}
\begin{figure}
\epsscale{0.3}
\plotone{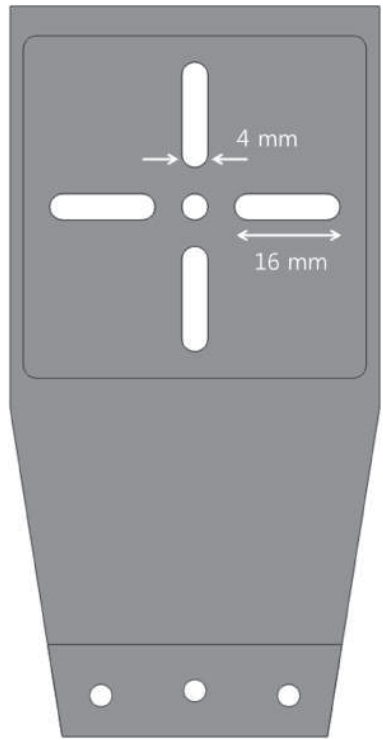}
\caption{3D modeling image of the initial filter mask for filter wheel calibration. This mask can be easily replaced with a science filter during observation.\label{IFM}}
\end{figure}

\begin{figure}
\epsscale{1.0}
\plotone{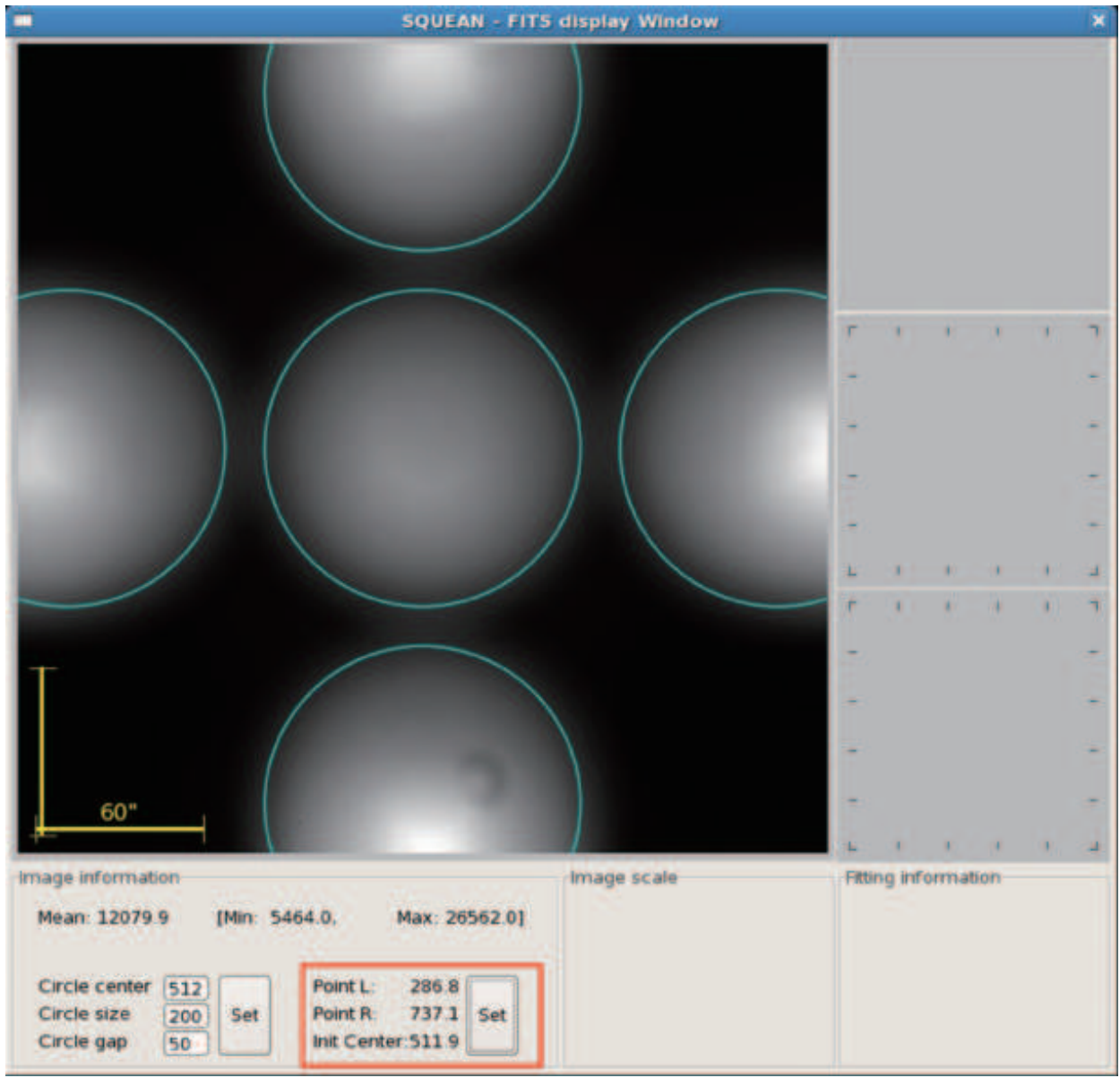}
\caption{Sample image of the quick look window for calibration. The values of Point L, Point R, and Init Center in the red box correspond to left, right edge points, and the center of filter, respectively.\label{Flat}}
\end{figure}

\begin{figure}
\plotone{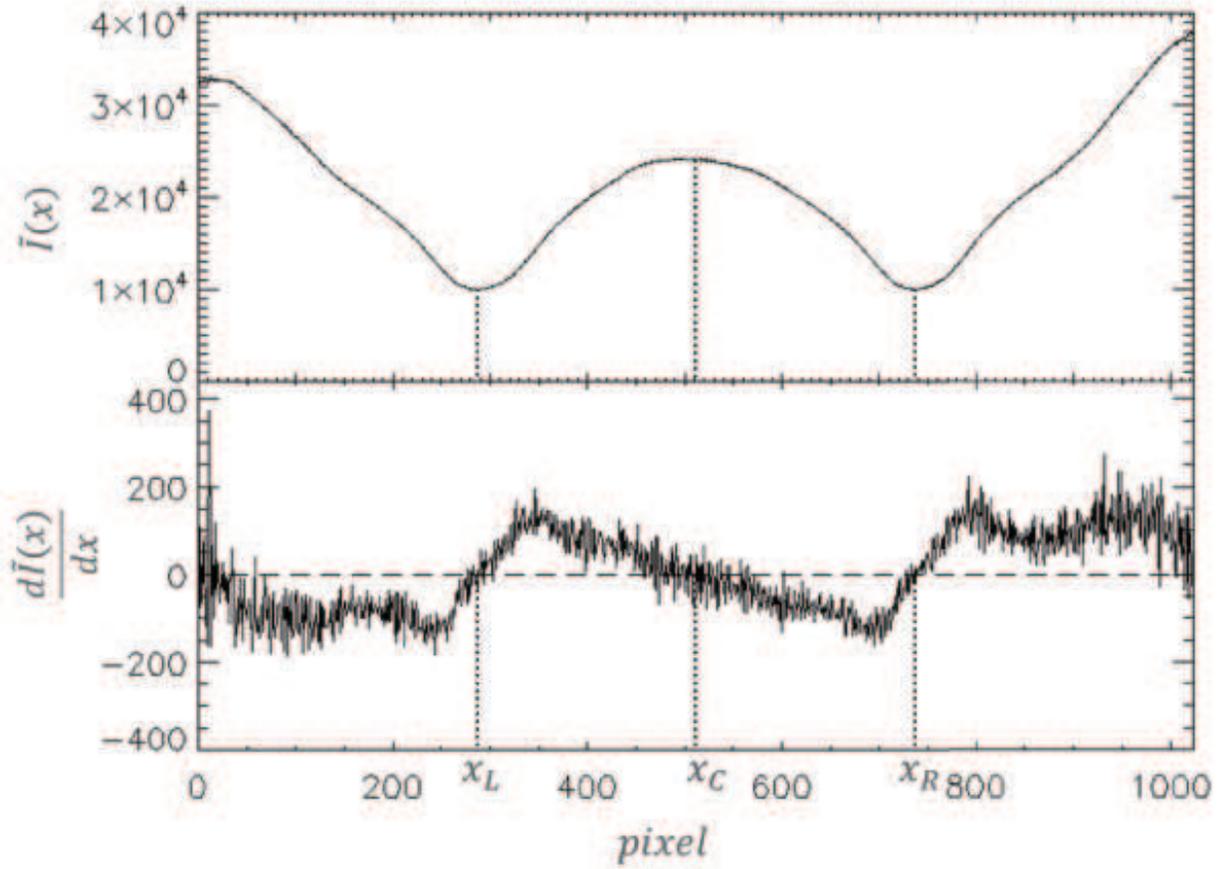}
\caption{Plots of median pixel values (upper) and the first derivative (lower) of the central 200 lines. The dotted lines indicate the derived edge positions ($x_L$ and $x_R$) and the center ($x_C$) of the filter.\label{Plot}}
\end{figure}

Figure \ref{IFM} shows the initial filter mask (IFM) that is used to test the stability of the filter wheel movement.
IFM also allows an observer to calibrate the filter wheel with high accuracy.
Figure \ref{Flat} shows a sample dome flat image taken through IFM in a quick look window.
The bright spots occur by the patterns on IFM, and the central circle should appear on the image center if the filter wheel operates correctly.
The pixel value at the pixel $(x,y)$ is defined as $I(x,y)$, and $\tilde{I}(x)$ is the median value of the central 200 lines.
We defined the edge positions on the left ($x_L$) and the right ($x_R$) as follows:
\begin{equation}\label{Edge}
\frac{d\tilde{I}(x_L)}{dx} = \frac{d\tilde{I}(x_R)}{dx} = 0,
\end{equation}
We applied a least square solution to estimate the solutions in Equation \ref{Edge}, and the center of the filter ($x_C$) is calculated by Equation \ref{Center}.
\begin{equation}\label{Center}
x_C = \frac{x_L + x_R}{2}
\end{equation}
Figure \ref{Plot} shows the $\tilde{I}(x)$ and its first derivative.
We can also check the center of the filter for the y-direction, which represents the manufacturing and assembly accuracy of the filter wheel.

We also checked the stability of the filter wheel in the laboratory, as well as during an engineering run of SQUEAN on-site.
The result shows that operation of the filter wheel is very stable and the shift in the filter wheel position after the wheel movement is unmeasurable ($\ll$ 1 pixel).

\section{Sensitivity}
\begin{figure}
\plotone{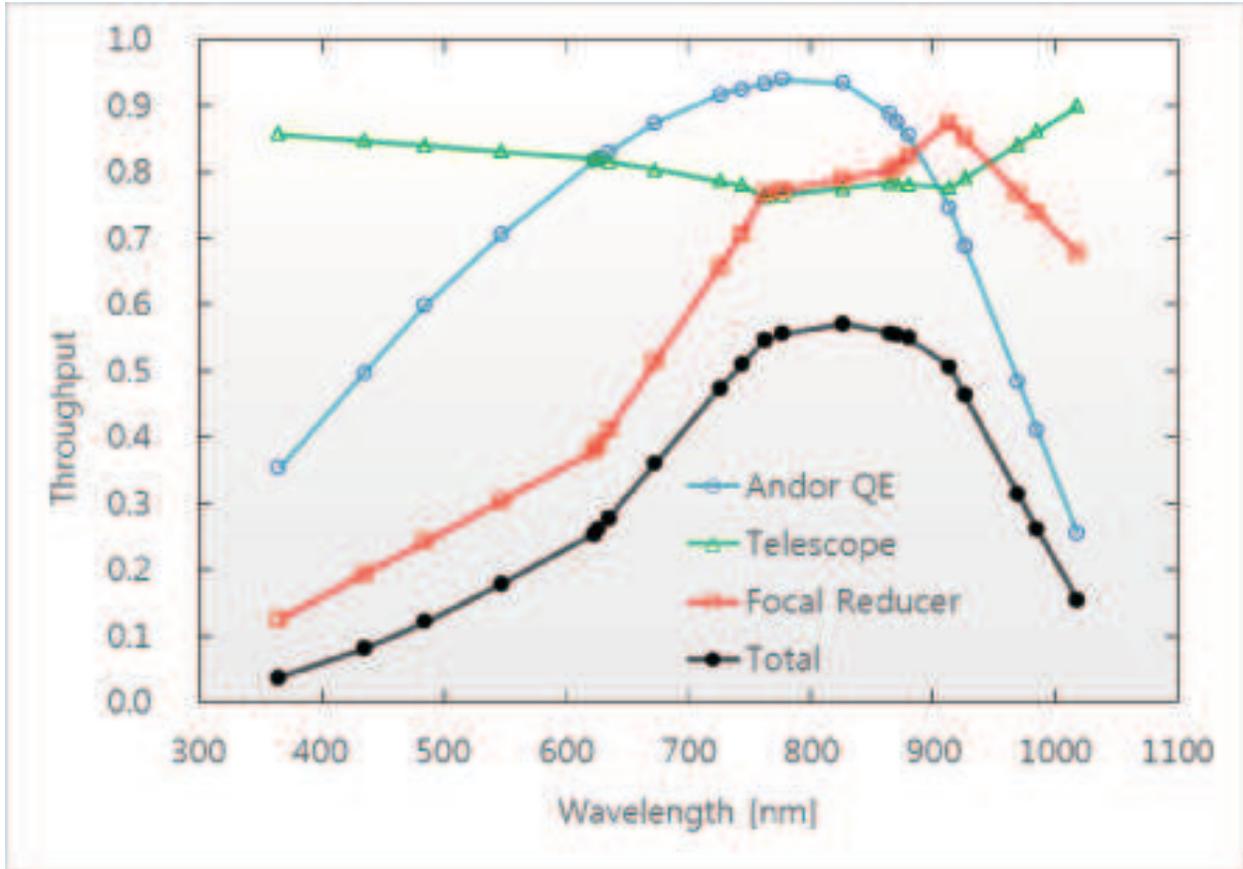}
\caption{Throughput of various components of SQUEAN. The blue open circles indicates the QE of the Andor CCD camera. The green triangles and the red squares represent throughputs of the telescope and the focal reducer, respectively. The black filled points show a total throughput of SQUEAN.\label{Throughput}}
\end{figure}

Figure \ref{Throughput} shows throughput curves for this system.
We measured the throughput of the focal reducer by comparing dome-flat images with and without the focal reducer.
The reflectivity of the telescope mirrors and the CCD quantum efficiency (QE) come from the data presented in Table 5 of \citet{Park2012}.
The total throughput, ${\tau}_{total}$, is defined as follows:
\begin{equation}\label{Total}
{\tau}_{total} = {\tau}_{tele} {\tau}_{fr} {\tau}_{qe},
\end{equation}
where ${\tau}_{tele}$ is the reflectivity of the telescope, ${\tau}_{fr}$ is the throughput of the focal reducer, and ${\tau}_{qe}$ is the QE of the CCD.
We found that the total throughput of SQUEAN is 0.4 -- 0.6 in the wavelength range of 700 nm -- 900 nm, while it is less than $\sim$0.2 for  $\lambda$ $<$ 600 nm and $\sim$0.3 for $\lambda$ $>$ 950 nm.
Based on the estimated throughput of the system, we calculated the expected limiting magnitude of each filter to compare with the limiting magnitude from the standard star observations.
The limiting magnitude is defined as the 5-$\sigma$ detection limit of a point source under a 1-hour total integration time and 1.$\arcsec$0 seeing condition.
We assumed that the sky background noise limit was the dominant noise source and estimated the limiting magnitude using Equation \ref{Limit} from \citet{Mc2008}.
The sky background data were obtained on 2015 May on a bright night.
\begin{equation}\label{Limit}
m_{lim} = 2.5\log{\biggl[\frac{\tau_{total}\tau_{filter}\lambda\Delta\lambda A_{tel}f_{aper}F_{\lambda}(0)}{hcg}\biggr]} - 2.5\log{\biggl[\frac{1}{g}\frac{S}{N}\sqrt{\frac{n_{pix}B}{T}}\biggr]},
\end{equation}
where, ${\tau}_{total}$ is the total throughput, ${\tau}_{filter}$ is the transmission of filter, $A_{tel}$ $[cm^2]$ is the telescope aperture, $f_{aper}$ is the fraction of flux of a point source that is contained within a given aperture, $F_{\lambda}(0)$ $[Wcm^{-2}\micron^{-1}]$ is the flux from the zeroth magnitude standard star of the AB magnitude system \citep{Oke1983}, $h$ is Planck's constant, $c$ is the speed of light, $g$ $[electrons/DN/sec]$ is the gain, $B$ $[DN/sec/pixel]$ is the average sky background, $S/N$ is the signal to noise ratio, $n_{pix}$ is the number of pixels covered by a circular aperture with a seeing size diameter of 1.$\arcsec$0, and $T$ is the total exposure time.
For a point spread function with a Gaussian profile, we adopted $f_{aper} = 0.76$ for an aperture diameter of FWHM.

We also directly measured the limiting magnitude using the data that were obtained from 2015 February and 2016 February.
We derived limiting magnitudes of 23.3 - 25.3 AB mag during dark times for a 1-hour total integration time with a 5-$\sigma$ detection limit under 1.$\arcsec$0 seeing condition.
Here, the detection limits were determined from the sky background sigma, within an aperture with a diameter of the FWHM for a point source.
Aperture correction was applied for the signal loss due to the use of a finite size aperture.

In Table 1, we list the limiting magnitudes.
Overall, the limiting magnitudes from different methods roughly agree with not only each other, but also the values presented in \citet{Jeon2016} and \citet{Park2012}.

\section{Summary}
In order to perform medium-band observation of objects in NIR, we developed SQUEAN, an SED camera system that is an upgraded version of CQUEAN and is installed at the Cassegrain focus of the 2.1 m telescope at the McDonald Observatory, USA.
SQUEAN consists of  an on-axis focal plane camera module, an auto-guiding system, and mechanical supporting structures.
SQUEAN also contains its in-house GUI software packages.
The primary improvement from CQUEAN to SQUEAN is the large filter wheel that can house twenty filters and its new software packages.
We employed eleven medium-band filters for medium-band photometry, which can be an efficient selection strategy to find high-z quasars compared to broad-band photometry.
In addition to the medium-band filters, the filter wheel currently includes broad-band filters, i.e., CQUEAN filters and Johnson-Cousin $BVRI$ filters.
The filter wheel has a flexibility of housing filters up to 86 mm $\times$ 86 mm, thanks to the usage of filter cartridges.
A finite element analysis of the filter wheel shows that the flexure of the filter wheel is at most 32 $\micron$, which is much smaller than the minimum space between the carousel and the outer box.
The filter wheel control package (KFC82) was newly developed, and the main observation software package (SMOP) and auto-guiding package (KAP82) were revised.
The filter wheel check with the initial filter mask shows that the filter position error is much less than 1 pixel during one night observation.
We derive the 5-$\sigma$ detection limits of 23.3 - 25.3 AB mag with a 1-hour total integration time, making it possible to observe faint, distant objects such as high redshift quasars.

We expect that SQUEAN will find  a wide range of applications in astronomical research.
Examples include the identification of high redshift quasars and brown dwarfs \citep{Jeon2016}, improvement of photometric redshifts of distant galaxies, reverberation mapping of active galactic nuclei, and studies of young stellar objects.

\section*{Acknowledgement}
This work was supported by a National Research Foundation of Korea (NRFK) Grant, No. 2008-0060544, and partially supported by NRF-2014M1A3A3A02034810, funded by the Ministry of Science, ICT and Future Planning (MSIP) of Korea.
S. Kim, H. I. Lee, and W. Park were partially supported by the BK21 plus program through a NRF funded by the Ministry of Education of Korea.
M. Hyun acknowledges the support from Global Ph.D. Fellowship Program through the National Research Foundation of Korea (NRF) funded by the Ministry of Education (NRF-2013H1A2A1033110).
This paper made use of the data obtained with a 2.1 m telescope at the McDonald Observatory, TX, USA.
We appreciate the staff of the McDonald observatory, including David Doss, Brian Roman, Kevin Meyer, Coyne A. Gibson, and John Kuehne, for their support during the installation of SQUEAN, and the CEOU members for obtaining some of the SQUEAN data that were used in this paper.

{}

\begin{deluxetable}{cccccc}
\tabletypesize{\scriptsize}
\tablecaption{SQUEAN Sensitivity}
\tablewidth{0pc}
\tablehead{
\multicolumn{3}{c}{Filter} & \multicolumn{3}{c}{Limiting Magnitude}\\
\multicolumn{3}{c}{Characteristics} & \colhead{Expected\tablenotemark{a}} &  \multicolumn{2}{c}{Measured}\\
\colhead{Name} & \colhead{${\lambda}_{effective}$ (nm)} & \colhead{FWHM (nm)} & \colhead{} & \colhead{Dark\tablenotemark{b}} & \colhead{Bright\tablenotemark{c}}
}
\startdata
B	&	435 	&	95	&	23.8 	&	-	&	23.6 	\\
V	&	548 	&	103	&	24.3 	&	-	&	24.0 	\\
R	&	635 	&	157	&	24.7 	&	25.3 	&	24.5 	\\
I	&	880 	&	166	&	24.9 	&	25.2 	&	24.7 	\\
r	&	623 	&	134	&	24.6 	&	25.2 	&	24.3 	\\
i	&	763 	&	145	&	25.2 	&	25.3 	&	24.7 	\\
is	&	744 	&	119	&	24.9 	&	-	&	-	\\
iz	&	865 	&	158	&	25.1 	&	-	&	-	\\
z	&	913 	&	148	&	25.2 	&	24.5 	&	-	\\
Y	&	985 	&	120	&	24.3 	&	-	&	-	\\
625	&	627 	&	50	&	23.9 	&	24.6 	&	23.4 	\\
675	&	673 	&	48	&	24.0 	&	25.0 	&	23.8 	\\
725	&	726 	&	49	&	24.1 	&	25.0 	&	24.0 	\\
775	&	777 	&	51	&	24.3 	&	24.6 	&	23.9 	\\
825	&	827 	&	47	&	24.2 	&	24.6 	&	23.9 	\\
875	&	870 	&	49	&	24.1 	&	24.2 	&	23.6 	\\
925s	&	926 	&	52	&	24.0 	&	23.8 	&	-	\\
925n	&	926 	&	26	&	23.7 	&	-	&	-	\\
975	&	969 	&	49	&	23.6 	&	23.3 	&	-	\\
1025	&	1018 	&	47	&	23.1 	&	-	&	-	\\
\enddata
\tablenotetext{\dagger}{With a 1-hour integration time and a 5-$\sigma$ detection limit under a 1.$\arcsec$0 seeing condition; the units of limiting magnitudes are AB mag.}
\tablenotetext{a}{Expected limiting magnitude that was estimated using background limit.}
\tablenotetext{b}{Derived using observation of a standard star on 2016 February.}
\tablenotetext{c}{Measured using observation of a standard star on 2015 February.}
\end{deluxetable}

\end{document}